# Time Slotted Scheduling Scheme for Multi-hop Concurrent Transmission in WPANs with Directional Antenna

Muhammad Bilal, Moonsoo Kang, Sayed Chhattan Shah and Shin-Gak Kang

To fulfill the requirements of high-speed, short-range wireless multimedia applications, millimeter-wave wireless personal area networks (WPANs) with directional antennas are gaining increased interest. Due to the use of directional antennas and mmWave communications, the probability of non-interfering transmissions increases in a localized region. The network throughput can immensely increase by the concurrent time allocation of non-interfering transmissions. The problem of finding optimum time allocation for concurrent transmissions is an NP-hard. In the literature, few "sub optimum concurrent time slot allocation" schemes have been proposed. In this paper, we propose two enhanced versions of previously proposed Multihop Concurrent Transmission (MHCT) scheme. To increase the network capacity, these schemes efficiently use the free holes in the time allocation map of MHCT scheme and make it more compact.

Keywords: High-speed WPANs, mmWave communication, Concurrent Transmission, Scheduling, Optimization

Manuscript received July 15, 2013; revised Oct. 29, 2013; accepted Nov. 5, 2013. This work was conducted as a part of the master's degree program of Muhammad Bilal at Chosun University under the supervision of Dr. Moonsoo Kang. The enhanced research was supported by the ICT Standardization program of MSIP (Ministry of Science, ICT & Future Planning)
Muhammad Bilal (phone: +82428606424, email: mbilal@etri.re.kr) is with Media application standard research laboratory, ETRI, Daejeon Rep. of Korea, Moonsoo Kang (email: mskang@chosun.ac.kr) Assistant professor Chosun University, Gwangju, Rep. of Korea, Sayed Chhattan Shah (email:shah@etri.re.kr) is with ETRI, Daejeon Rep. of Korea and Shin-Gak kang (email: sgkang@etri.re.kr) is with Media application standard research laboratory, ETRI, Daejeon Rep. of Korea

## I. Introduction

To achieve high-speed connectivity for short-range wireless multimedia applications such as high-definition TVs, kiosk file servers, and HD audio, millimeter-wave wireless personal area networks with directional antennas are gaining increased interest. As the physical layer in standardizations and specifications such as IEEE 802.15.3 [1] and IEEE 802.11 VHT [2], mmWave communication with directional antennas at the unlicensed 60 GHz band was adopted because this spectrum can achieve multi-gigabit link speed (conceivably 3.5 Gbps) [3]. Due to the important characteristics of a high propagation loss over distance in mmWave communications, spatial usability has become very high [4], [5]. Since the overlapped transmission area of directional antennas is smaller than that of Omni-directional antennas, further spatial reusability can be achieved using directional antennas.

Furthermore, in the high-frequency band, reflection is more dominant than diffraction at the receivers. In addition, the performance at 60 GHz is highly dependent upon the obstructions between the source and destination nodes. Therefore, achieving a high data rate while maintaining the line of sight (LOS) is a key factor [6], [7]. To maintain as short a distance and LOS between a transmitter and receiver as possible, a relay node is introduced [8], which helps to achieve higher data rates between a transmitter and receiver. Without a relay, the transmission will be interrupted, and the connectivity will experience a serious link outage from moving obstacles.

The network throughput enhancement schemes for mmWave WPANs have been discussed in the literature [8-12]. [8] Proposed an architecture mmWave WPAN, where a relay node is selected when the LOS link between source and destination is blocked by moving obstacles. Without relay, the transmission will be interrupted and the connectivity will experience a serious link outage by moving obstacles. In [9] and [10] authors have developed an exclusive region (ER) based resource management scheme and analytically derived the optimal ER sizes to explore the spatial multiplexing gain of mmWave WPANs with directional antenna. In [11] author enabled the concurrent transmissions of noninterfering transmissions. However, [11] is limited in terms of single hop or minimum hops relay for data transmission. Thus, [12] proposed a multi-hop concurrent transmission (MHCT) scheduling algorithm. In this paper, we analyzed MHCT and found that its time allocation map of non-interfering concurrent transmissions is not fully compact, and that free holes exist in the time allocation map.. Further improvement in network throughput is possible by utilizing these holes (by considering inter-group collisions). Hence, we extended the MHCT [12] and proposed two new schemes: 1) enhanced multi-hop concurrent transmission with expandable group size (EMHCT-E) and 2) enhanced multi-hop concurrent transmission with fixed group size (EMHCT-F). We also introduced more efficient conditions for selecting relay nodes, and modified the priority scheme of "transmission selection" to prevent a starvation. In addition, we introduced a "concurrency gain" to find the theoretical bound of the network throughput using a water-filling algorithm. Finally, we made a "fairness" comparison of the proposed schemes using Jain's Fairness Index.

The remainder of this paper is organized as follows. In section II, we formally discuss a network system model. In section III, we describe our analysis of the MHCT and present the proposed algorithms (EMHCT-E/F). In section IV, the simulation parameters and performance metrics are defined, and extensive simulation results are presented to compare the proposed algorithm with MHCT and water-filling theoretical bound. Finally, we provide some concluding remarks regarding this paper in section V.

## II. System Models

We consider an indoor single-hop WPAN with 50 wireless terminal nodes and a piconet controller (PNC). Each wireless terminal node is equipped with multiple steerable directional antennas. As the network size in WPAN is small and has low levels of mobility, we assume that during a random access period, PNC can receive the location information of each node.

1. Notations

The following notations are used throughout the rest of this paper.
- $R_i$ = $i$-th transmission request.
- $n(i)$ = time slot requirement by $i$-th transmission request.
- $h_k^{R_i}$ = $k$-th hop of $i$-th transmission request.
- $n(I,K)$ = time slot requirement by $k$-th hop of $i$-th transmission request.
- $d(i,j)$ = distance between $i$–th and $j$-th nodes.
- $w(i,j)$ = link weight between $i$–th and $j$-th nodes.
- $F(j)$ = workload of $j$-th node.
- $G_i$ = $i$-th group of concurrent hop transmissions.
- $n(G_i)$ = time slot requirement by $i$-th group $G_i$.
- $MAXSLOTS$ = number time slots in a suprframe.
- $N_{slots}$ = available slots in a superframe.
- PNC = piconet controller.

2. Antenna model

We considered an ideal "flat-top" antenna model for directional antenna [13]

$$G(\emptyset) = \begin{cases} \frac{1}{N}\frac{\sin(\frac{N}{2}\pi\sin\emptyset)}{\sin(\frac{1}{2}\pi\sin\emptyset)}, & |\emptyset| \leq \frac{\Delta\emptyset}{2} \\ \ll 1, & otherwise \end{cases} \quad (1)$$

where $\Delta\varphi = 2\pi/N$ is the antenna beamwidth when every node is equipped with an antenna with $N$ beams, each of which spans an angle of $2\pi/N$ radians. Thus, if a receiver is directed within the antenna beamwidth of the transmitter, i.e., ($|\varphi|\leq\Delta\varphi/2$), the antenna gain of the transmitters and receivers is $G_t = G_r = \frac{1}{N}\frac{\sin(\frac{N}{2}\pi\sin\emptyset)}{\sin(\frac{1}{2}\pi\sin\emptyset)}$ dBi [13], while $G_t = G_r << 1$ if the node resides outside of the transmitter beamwidth. In addition, in LOS room case, the received power is mainly a directed wave [14]. Hence, in the antenna model discussed above, the interference outside the antenna beam is small enough to allow a concurrent transmission, while inside the beam width is large enough to block another transmission.

3. mmWave communication rate and time slot calculation

An indoor environment is less dynamic compared to an outdoor environment, and thus we can assume that the channel conditions remain almost static for the time duration of a superframe. In IEEE 802.15.3, the throughput mainly depends on the scheduling scheme rather than transmission power [15]. We can assume that all nodes can transmit with constant maximum power (P). The achievable data rate according to Shannon's theory is given by;

$$R = W \log_2[1 + \frac{P_r}{(N_o + \sum P_r^i)Wr^n}] \quad (2)$$

where $W$ is the system bandwidth; $N_0$ and $I$ are the one-side power spectral density with white Gaussian noise and interference, respectively; $P_r$ is the received signal power; $P_r^i$ is the received signal power from interfering transmission; $G_r$ and $G_t$ are the antenna gain of the receiver and transmitter, respectively; $\lambda$ is the wavelength; $r$ is the transmission distance between the transmitter and receiver; and $n$ is the path loss exponent whose value is usually between 2 and 6 for an indoor environment [16]. According to Friis free space equation,

$$P_r = \frac{P_t \lambda^2 G_t G_r}{(4\pi)^2 r^n} \quad (3)$$

Form (1) and (2) we get,

$$R = W \log_2[1 + \frac{P_t G_t G_r \lambda^2}{16\pi^2(N_o + \sum P'_t G'_t G_r \lambda^2/16\pi^2 r^n)Wr^n}] \quad (4)$$

Where $P'_t$ and $G'_t$ are the transmission power and antenna gain of other transmitting nodes, respectively.

According to antenna model $G'_t \ll 1$ for non-interfering transmissions. Therefore the interfering term ($\sum P'_t G'_t G_r \lambda^2 / 16\pi^2 r^n$) in (4) is insignificant for concurrent scheduling of non-interfering transmissions.

Once we have $R$, the time slot $n(I,J)$ requirement for a transmission request from $h_{k-1}^{Rm}$ to $h_k^{Rm}$ with a P Mb/s data payload is given below:

$$n(I,J) = \frac{P/R}{t_{ts}} \quad (5)$$

where $t_{ts}$ is a single time slot duration.

### 4. Directional MAC structure

The IEEE 802.15.3 superframe structure in Fig. 2 is used for directional MAC. Directional MAC applies the same logic as MAC, except it gives access control on a per antenna basis.

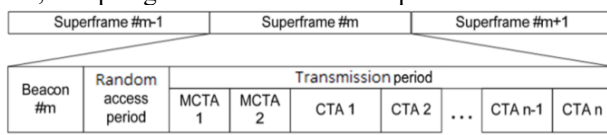

Fig. 1. IEEE 802.15.3 MAC.

A super frame is composed of a beacon period, random access period, and transmission period. During a beacon period, the PNC broadcasts the synchronization and scheduling information. The scheduling information includes the start time and duration of the transmission period, and the direction of the steering beam. During a random access period, nodes willing to transmit data send transmission requests to the PNC. The transmission request includes the topology information used to determine the transmitter's antenna direction and the node's work load. During the transmission period, only scheduled nodes are allowed to send their data for the duration of the allocated time slots i-e Channel Time Allocation (CTAs).

## III. Time Slot Allocation for Concurrent Transmission

The time slot allocation and scheduling of a concurrent transmission can be considered an optimization of the packing problem, where each transmission request can be considered an item having a variable width with interfering and conflicting dimensions. Let $[R_i, n(i)]$ denote each transmission request arriving at the PNC during the random access period along with its arrival order. Then, $R_i$ will be transformed into multihop transmissions (to by overcome the high path loss factor of mmWave communication) using the following hop selection metric used in [12].

$$w(i,j) = \frac{d^2(i,j)}{\overline{D^2}} + \frac{F(j)}{\overline{F}} \quad (6)$$

Let $\{[h_k^{Ri}, n(I,k)], k = 1, 2, …, m \text{ and } i = 1, 2, …, N\}$ denote an ordered sequence representing a multihop transmission for $R_i$, and $m$ be a number between 1 and $n(n-1)/2$, where $n$ is the number of nodes. For a $h_j^{Ri}$ hop transmission, $n(I,J)$ represents the required number of time slots. For example, the ordered sequence for $[R_1, n(1)]$ is $\{[h_1^{R1}, n(1,1)], [h_2^{R1}, n(1,2)], …, [h_k^{R1}, n(I,k)],\}$. The optimization problem of a time slot allocation within a superframe for a concurrent transmission can then be formulated as follows:

P1: $\max \sum_{i=1}^{n_f} R_i \quad (7)$

s.t $\sum_{i=1}^{n_f} \sum_{k=1}^{n_h^i} n(i,k) < MAXSLOTS$

$\forall \ h_k^{Ri} \ \{i = 1,2…,n_f\}, \{k = 1,2…,n_h^i\} \quad (8)$

To solve this problem, the optimum result leads to NP-hard [17]. Therefore, instead of solving the problem for the optimum result, a practical sub-optimum result is possible using a Heuristic approach.

### 1. Time slot allocation process in MHCT

In MHCT scheme once the direct transmissions are converted into multihop transmissions, PNC *sorts* the hop transmission requests in decreasing order according to the number of time slot requirements, $n(I,J)$. The selection of this priority scheme is made because the slow links will take advantage of the high priority, which leads to a relatively good fairness.

After sequencing the hop transmission requests, PNC checks for the concurrent hop transmissions in *hop sequence* order of each transmission request, and finally forms group $G_i$ of hops, which can be transmitted concurrently. The main consideration of MHCT is to identify and group all non-interfering hop transmissions into a group such that the condition of the coexistence of two or more hop transmissions of the same collision property (conflicting and/or interfering $h_k^{Ri}$) in the same group should not occur. This process continues until one of the following conditions is satisfied.

a) $\sum_{i=1}^{n_g} n(G_i) < MAXSLOT \quad \forall \ G_i\{i = 1,2 \dots, n_g\}$

b) All requests are scheduled.

When the next hop of $R_i$ is to be schedule in the next superframe, PNC recalculates multi-hop transmissions from the source node of the current hop transmission to the final destination node. After finishing the scheduling of the last transmission request, we may obtain the transmission scheduling map, such as in Fig. 2. Each group $G_i$ comprises hop transmissions which can transmit concurrently. The size of group $G_i$ is determined by the hop transmission $h_j^{Ri}$ with highest number of time slots $n(i,j)$ requirement e.g, first group $G_1$ has size $n(G_1) = n(3,1)$ i.e PNC allocates $n(3,1)$ number of slots for all hop transmissions $h_j^{Ri} \in G_1$ and so on.

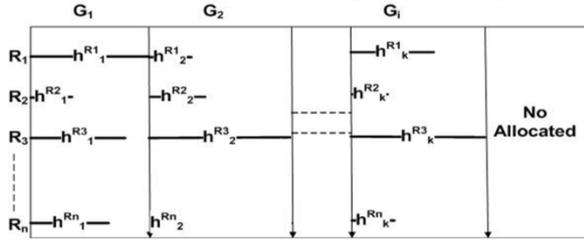

Fig. 2. Time slot allocation map of *MHCT*

The total consumption of time slots by the MHCT allocation map is then given below:

$\sum_{i=1}^{n_g} n(G_i) \quad \forall \ G_i\{i = 1,2 \dots, n_g\}$ (9)

Higher bandwidth efficiency will be achieved with a smaller value of (9).

2. Weaknesses and improvement of MHCT

*A. Imperfection in time allocation of MHCT*

In the time allocation process, MHCT does not consider inter-collisions among concurrent groups. The hop transmissions within a group guaranteed to have no collisions but hop transmissions between groups are not checked to determine if they are interfered with. By a span overlapping of groups, we can obtain a further compact mapping. In MHCT, conflicting/interfering transmissions cannot coexist in the group, even if they are not overlapping in time. Through scheduling the conflicting/interfering transmissions within a group in non-overlapping time, a further compact mapping can be achieved.

*B. Condition for multihop conversion*

After a mutihop conversion, the transmission graph becomes more complex, and few bottleneck links can emerge. Therefore, multihop conversion is not always beneficial. Furthermore removing the bottleneck at the beginning of each superframe will introduce significant complexity in the system. To reduce the complexity and obtain the full benefit of a hop conversion of each direct transmission, ($R_i$) is only converted into a multihop transmissions ($h_k^{Ri}$) if

$\sum_{i=1}^{n} \sum_{j=1}^{n_h^i} [h_j^{Ri}, n(i,j)] < \sum_{i=1}^{n} [R_i, n(i)].$ (10)

*C. Starvation problem in the priority scheme*

Under MHCT priority scheme if the nodes are very unevenly distributed or the number of transmission requests is very high, a starvation may occur. To resolve the starvation problem, we use the following aging policy to increase the priority of a suffering transmission request by 25% on each miss. It makes certain that a transmission request will get highest priority after four misses:

$P_{new} = 0.25 \ M_{count} \ P_{prev}$ (11)

where $P_{new}$ is a new priority, $P_{prev}$ is a previous priority, and $M_{count}$ is a counter incremented by "1" on each miss.

3. Time slot allocation process in EMHCT-F/E

To overcome the shortcomings of MHCT discussed in the previous section, we proposed two schemes, Enhanced Multihop Concurrent Transmission-Fixed Group (EMHCT-F) and Enhanced Multihop Concurrent Transmission-Expandable Group (EMHCT-E). The main objective is the identification and grouping of hop transmissions such that two or more conflicting/interfering hop transmissions can coexist in the same group if they follow subsequent conditions.

a) Conflicting and interfering transmissions should not overlap in time when they are in the same group.

b) They should follow *the hop sequence order* of each transmission.

c) For EMHCT-F, the time slots requirement $n(I,J)$ should satisfy condition '1', and for EMHCT-E, the time slot requirement $n(I,J)$ should satisfy condition '2'.

1. $n(I,J)$ should be less than or equal to the difference of the largest time slot requirement of conflicting hop transmissions and time slot requirement of group $n(G)$.

   - $n(I,J) <= n(G) - max[n_c(I,J)]$

2. $n(I,J)$ should be less than or equal to the sum of the remaining time slots in the superframe and the time slot requirement of group $n(G)$.

   - $n(I,J) <= N_{slots} + n(G) - max[n_c(I,J)]$, where $N_{slots}$ is the available time slots in a superframe.

**Algorithm 1  EMHCT-F**
1. BEGIN:
2. PNC receives a request $h_j^{Ri}$ for n (I, J) time lots
3. Sort all hops according priority policy
4. Start a new group G(k) with n(k) = max[n(I,J)]
5. while $N_{slots}$ ≥min[n(I,J)] or all $h_j^{Ri}$ scheduled
6.   if $h_j^{Ri}$ does not conflict with existing hops in G$_b$ then
7.     if n(b)≥ n (I, J) then
8.       Update G$_b$ = G$_b$U{$h_j^{Ri}$};
9.       Sort all hops according to priority policy.
10.     else
11.       if $N_{slots}$ ≥min[n(I,J)] then
12.         Start a new group G(k) with n(k) = max[rest of n(I,J)]
13.       end if
14.     end if
15.   end if
16. end while
17. for all non-empty group (G$_i$! = Null) ,{i=1,2….b-1} do
18.   if $h_j^{Ri}$ is in conflict with few of existing hops in Gi
19.     G$_c$ = Identify conflicting hops,  where G$_c$ ⊆ G$_i$
20.     n(c) = Maximum n(I,J) in G$_c$
21.     for all $h_j^{Ri}$ in G$_c$ do
22.       if n (I, J) ≤ n(i) - n(c)
23.         Update G$_i$ = G$_i$ ∪ i;{ $h_j^{Ri}$ }, position at n(c);
24.       end if
25.     end for
26.   else
27.     Update G$_i$ = G$_i$ ∪ i;{ $h_j^{Ri}$ }
28.   end if
29. end for
30. if $N_{slots}$ ≥min[rest of n(I,J)]
31.   go to line 5
32. end if
33. END;

**Algorithm 2  EMHCT-E**
1. BEGIN:
2. PNC receives a request $h_j^{Ri}$ for n (I, J) time lots
3. Sort all hops according priority policy
4. Start a new group G(k) with n(k) = max[n(I,J)]
5. while $N_{slots}$ ≥min[n(I,J)] or all $h_j^{Ri}$ scheduled
6.   if $h_j^{Ri}$ does not conflict with existing hops in G$_b$ then
7.     if n(b)≥ n (I, J) then
8.       Update G$_b$ = G$_b$U{$h_j^{Ri}$};
9.       Sort all hops according priority policy.
10.     else
11.       if $N_{slots}$ ≥min[n(I,J)] then
12.         Start a new group G(k) with n(k) = max[rest of n(I,J)]
13.       end if
14.     end if
15.   end if
16. end while
17. for all non-empty group (G$_i$! = Null) ,{i=1,2….b-1} do
18.   if $h_j^{Ri}$ is in conflict with few of existing hops in Gi
19.     G$_c$ = Identify conflicting hops,  where G$_c$ ⊆ G$_i$
20.     n(c) = Maximum N(I,J) in G$_c$
21.     for all $h_j^{Ri}$ in G$_c$ do
22.       if n (I, J) ≤ $N_{slots}$ + n(i) - n(c)
23.         Update G$_i$ = G$_i$ ∪ i;{ $h_j^{Ri}$ }, position at n(c);
24.         if n(I, J) ≥ n(i)
25.           Update  n(i)=n(I, J);
26.         end if
27.       end if
28.     end for
29.   else
30.     Update G$_i$ = G$_i$ ∪ i;{ $h_j^{Ri}$ }
31.   end if
32. end for
33. if $N_{slots}$ ≥min[rest of n(I,J)]
34.   go to line 5
35. end if
36. END;

A. EMHCT-E/F algorithm

Algorithms 1 and 2 are the enhanced versions of MHCT, which consider inter-group and intragroup collisions to schedule hop transmission requests. A brief stepwise explanation is given below.

a) STEP 1: Execute MHCT with an improved mutihop conversion condition according to (10) and adjust the priorities using the improved version according to (11).

b) STEP 2: Start the span overlapping process from $G_2$ against $G_1$. After finishing the span overlapping between $G_2$ and $G_1$, apply the same procedure to $G_3$ against $G_2$ and so on. Start the span overlapping of $G_2$ from the first hop transmission of $G_2$. Check if this hop transmission or the span overlapping candidate causes a collision with the hop transmissions in $G_1$ one by one until meeting a hop transmission with a collision, or a hop transmission belonging to the same transmission request. If a span overlapping candidate finds a few collisions or hop transmissions from the same transmission request, EMHCT-F checks conditions a, b and c-1, whereas in the case of EMHCT-E, it checks conditions a, b and c-2.

c) STEP 3: If the specific hop transmission satisfying the above conditions is found through the lookup in STEP 2, the allocated time slots of the span overlapping candidate move back to back at the end of the hop transmission before the specific hop transmission. The same procedure in STEP 2 is then performed for the next hop transmission of $G_2$. After finishing a span overlapping of all hop transmissions in $G_2$, the hop transmissions of $G_3$ start the span overlapping procedure described in STEP 2 and STEP 3. The span overlapping procedure will continue until finishing the span overlapping of the last group.

EMHCT-E and EMHCT-F both outperform MHCT for different beamwidths. However, the performance of *EMHCT-E* is better than EMHCT-F for a large beamwidth, whereas the EMHCT-F performance is better than EMHCT-F for a small beamwidth.

For a large antenna beamwidth, each transmission occupies a larger area with a large dimension of interference. Therefore, without altering the size of a group, it becomes difficult to place a new transmission request in pre-existing groups. The expansion of a group, which should satisfy condition c-2, increases the probability of placing the new transmission request in the existing groups. Hence, EMHCT-E has better results compared to EMHCT-F. For a smaller beamwidth, each transmission occupies a smaller area with a small dimension of interference. Therefore, without altering the size of a group, we can place a new transmission request in already existing groups, which should satisfy condition c-1. Hence, EMHCT-F has better results compared to EMHCT-E.

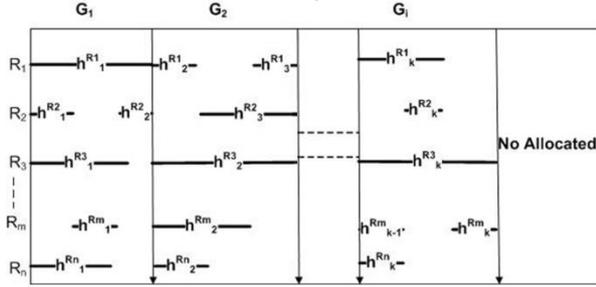

Fig. 3. Time slot allocation map of EMHCT-F/E

In a general pictorial form, EMHCT-E and EMHCT-F both provide the same time slot allocation map for concurrent transmissions, as given in Fig. 3. In EMHCT-F/E, each group holds more hop transmission requests compared to MHCT, and hence the number of groups will be reduced for the same number of hop transmission requests. The size of each group will also be nearly the same because the size of a group is determined based on the priority scheme, which is the same for MHCT and EMHCT-F/E. This implies the following inequality:

$$\sum_{i=1\ EMHCT-E/F}^{n_g} n(G_i) < \sum_{i=1\ MHCT}^{n_g} n(G_i) < \sum_{i=1}^{i} \sum_{k=1}^{n_h^i} [h_j^{Ri}, n(i,j)] < \sum_{i=1}^{i} [R_i, n(i)]$$

In Fig. 3, hop transmission $h_2^{R2}$ in $G_l$ has interference with $h_1^{Rm}$, and the previous hop transmission ($h_1^{R2}$) also already exists in $G_l$. Because $n(2,2)$ is less than $[n(1,1)-(n(m,1)+n(2,1))]$, $h_2^{R2}$ is placed in $G_l$, such that it does not overlap with $h_1^{Rm}$ and scheduled after $h_1^{R2}$. Similarly, $h_3^{R1}$ is placed in $G_2$ along with conflicting hop $h_2^{Rm}$ and the previous hop transmission ($h_2^{R1}$), as this satisfies all conditions necessary to avoid a collision during concurrent transmissions.

## IV. Simulation and Performance Evaluation

### 1. Simulation settings

Thirty nodes were randomly deployed in a room 16 x16 m in size. Each node has multiple antennas. The number of antennas depends on the beamwidth used.

$$No. of\ Antennas = \frac{360}{Beam\ width\ in\ degrees} \quad (12)$$

Table 1  Simulation parameters.

| Parameters | Value |
|---|---|
| Bandwidth | 7000MHz |
| Transmission Power | 0.1 mW |
| Antenna Gain | 12dBi |
| Background noise | -134dBm/MHz |
| Path loss exponent | 3~6 |
| Antennas | 18,8,4,2 |
| Bandwidth | 7000MHz |

We considered 7 GHz of bandwidth for our simulations (IEEE 802.15.3c defines the use of 9 GHz of bandwidth (57 to 66 GHz); however, in Korea, the USA, and Japan, 7 GHz of bandwidth is available). The rest of the parameters were selected according to [12] and [15].

The simulation was performed using different amounts of data, with each data traffic flow varying from 50 to 350 mb. The nodes were randomly deployed and simulated for different numbers of active traffic flows. The number of active traffic flows varied from 1 to 50, and for each simulation run, traffic flow pair selection was also conducted randomly using ten different seed values. For each beamwidth selection, a total of 700 simulations were carried out; the results were taken by averaging all of the simulation runs for each beamwidth selection. In our simulation, the computational cost of the antenna selection was not taken as a parameter. If we consider the computational cost of the antenna selection, there will be an upper bound to the number of antennas required to obtain the highest throughput.

### 2. Performance parameters

To compare and determine the performance of our algorithm, we considered the following performance parameters.
  a) Throughput: We calculated the network throughput (the total volume of data traffic through the network) to check the bandwidth efficiency achievement across the network using the proposed algorithms.
  b) Fairness: A greedy network system, which is designed to achieve a higher network throughput, usually leads to unfair resource sharing. Hence, from a user perspective, few users (with good channel conditions) receive a very high data rate, and other users suffer from an extreme low data rate. Our capacity gaining algorithm takes care of this problem and provides high throughput with acceptable fairness. We used Jain's fairness index to measure the fairness of the proposed systems.

c) Concurrency gain: Concurrency gain is defined as a ratio between the network throughput achieved by concurrent transmission and the corresponding network throughput by direct transmission or a ratio between time slots requirement by direct transmission to the time slots requirement by concurrent transmission.

The concurrency gain determines the aggregate improvement achieved by the concurrency compared to a direct transmission under same network configuration. The concurrency gain is given below.

$$\rho = \frac{n_d(i)}{n_c(i)} = \frac{R_i^c}{R_i^d} \quad (13)$$

where $n_d(i)$ is time slot requirement for a direct transmission, $n_c(i)$ is time slot requirement for a concurrent transmission, $R_i^c$ is the data rate achieved for concurrent scheduling, and $R_i^d$ is the data rate achieved for a direct transmission.

### 3. Optimality comparison (water filling)

A water-filling solution is a well-known algorithm used to provide the theoretical bound for the capacity gaining constrained optimization problem. A generalized and simple algorithm for the water-filling problem is presented in [18]. The solution is provided under a power constraint with the objective of an optimization of the power transmission within a single frame. However, with minor changes and assumptions, the solution can be used for the theoretical bound for an optimization of the time allocation process. If we reconsider our objective function of the optimization problem P1 (defined in section-III) and redefine according to the form of the constraint optimization problem discussed in [18]. Then we can obtain the water-filing result to calculate the optimum capacity gain using algorithm 3. The objective of the following problem P2 is log concave, which ensures proportional fairness (PF) with optimum throughput [19].

P2: $\max \sum_{i=1}^{n_f} \log(1 + R_i \rho_i)$

s.t $\sum_{i=1}^{n_f} \sum_{k=1}^{n_h^i} n(i,k) < MAXSLOTS \quad \forall h_k^{Ri}$
where, $\{i = 1,2 \dots, n_f\}, \{k = 1,2 \dots, n_h^i\}$ (14)
Given by
$n(i,k) = (\mu - \rho_i^{-1})^+$
where, $\{ i = 1,2 \dots, n_f\}, \{k = 1,2 \dots, n_h^i\}$ (15)
where $(\propto)^+$ selects the maximum value for $n(i,k)$, $i*k$ are the total hop transmission requests $h_k^{Ri}$, and $\mu$ is the water level, which is chosen such that $\sum_{i=1}^{n_f} \sum_{k=1}^{n_h^i} n(i,k) = MAXSLOTS$. Once $\mu$ is selected, the depth from the water level depends upon the concurrency gain $(\rho)$. The value of $\mu$ deepens if the concurrency gain $(\rho)$ is high. This means that the hop transmission requests $h_k^{Ri}$ will obtain a higher data rate with a low water level.

Algorithm 3 provides a water-filling solution with the worst case complexity of $i*k$ iterations. In algorithm 3, constraint function $g$ satisfies the constraint condition, i.e., $g(\mu) = \sum_{i=1}^{n_f} \sum_{k=1}^{n_h^i} n(i,k)_\mu - MAXSLOTS$. This constraint function makes the value of $n(i,k)$ dependent upon water level $\mu$. In this way, the hop transmission requests $h_k^{Ri}$ with high concurrency gain $(\rho)$ receive a higher allocation of time resources.

---

**Algorithm 3 Water-filling solution**

Input: Set of concurrency gain $\{(\rho)\}$ and constraint function $g$.
Output: Numerical solution $\{n(i,k)\}$ and water level.
1. Set $\tilde{l} = i*k$, and sort the set $\{(\rho)\}$ such that $\rho_i$ are in decreasing order $\rho_i^{-1} > \rho_{i+1}^{-1}$ (define $\rho_l^{-1} = 0$)
2. If $\rho_{\tilde{l}} < \rho_{\tilde{l}+1}$ and $g(\rho_{\tilde{l}})$ then accept and go to step 3. Otherwise, reject form new one by setting $\tilde{l} = \tilde{l} - 1$ and go to step 2.
3. Find water level $\mu \in (\rho_{\tilde{l}}, \rho_{\tilde{l+1}}) | g(\mu) = 0$, obtain numerical solution as, $n(i,k) = (\mu - \rho_i^{-1})^+$ $\{i = 1,2 \dots, n_f\}, \{k = 1,2 \dots, n_h^i\}$
4. Undo sorting done at step 1 and finish.

---

This greedy yet proportional fair approach increases the overall throughput of the network to provide the optimum result.

### 4. Results

#### A. Beamwidth effect

The effect of the beamwidth on aggregate concurrent transmissions is significant. With a small antenna beamwidth, the chance of a concurrent transmission increases due to small coverage area per transmission; also the antenna gain of small beamwidth is high. Hence the network throughput increases. Figures 4 through 7 show the throughput comparison of MHCT, EMHCT-F, and EMHCT-E, with different beamwidth selections. In all cases, EMHCT-F and EMHCT-E perform better than MHCT, because both provide more compact time allocation map compare to MHCT. EMHCT-E provides a better throughput for large beamwidths (Fig. 6 and 7), but the increment in the performance with respect to a reduction of the beamwidth is slower as compared to EMHCT-F. Hence, EMHCT-F performs better for a beamwidth smaller than 45 deg (Fig. 4). For an antenna beamwidth of 45deg, EMHCT-E and EMHCT-F performance is almost same as shown in Fig. 5.

The reason for this behavior is obvious because EMHCT-F has a tendency to provide more opportunities for hop transmission requests $h_k^{Ri}$ with fewer time slot $n(i,k)$ requirement. For a small antenna beamwidth, the probability of an interference is reduced, and the number of hop transmission request $h_k^{Ri}$ with fewer time slot $n(i,k)$ requirement increases. Hence if the group size is fixed the hop transmission request $h_k^{Ri}$ with fewer time slot $n(i,k)$ requirements receives more opportunities to be scheduled by finding small rooms in previously created groups (without altering the size of group). Therefore, EMHCT-F outperforms EMHCT-E for a smaller antenna beamwidth as shown in Fig. 4.

In contrast, for a large antenna beamwidth, the probability of interference increases, and the number of hop transmission request $h_k^{Ri}$ with more time slot $n(i,k)$ requirement increases. Hence if the group size is fixed the hop transmission requests $h_k^{Ri}$ have fewer opportunities to be scheduled in a previously created fixed-sized group, which degrades the performance of EMHCT-F. For, EMHCT-E a previously created group size is expandable during the span overlapping of groups. Therefore, EMHCT-E has a tendency to provide more opportunities for hop transmission requests $h_k^{Ri}$ with higher time slot $n(i,k)$ requirements, and through a group expansion, the probability to schedule transmission requests $h_k^{Ri}$ increases, which leads to a better performance of EMHCT-E for a large antenna bandwidth as shown in Fig. 6 and 7.

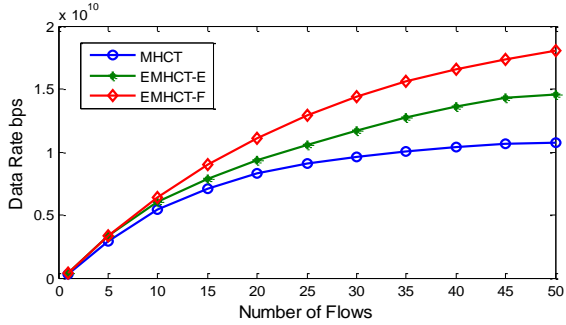

Fig. 4. Throughput for 20 deg beamwidths

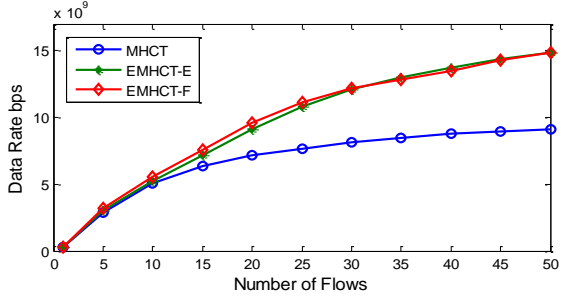

Fig. 5. Throughput for 45 deg beamwidths

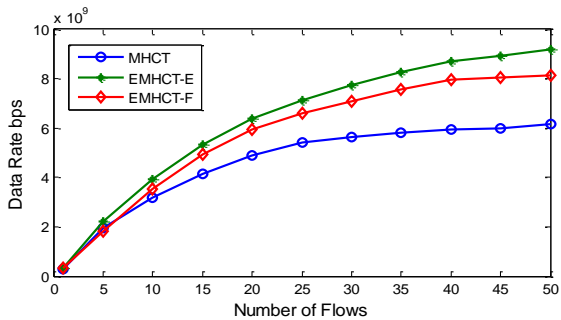

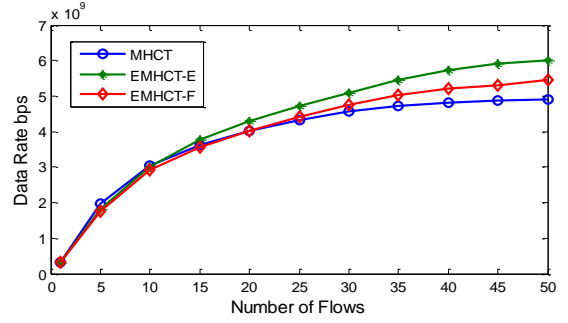

Fig. 6. Throughput for 90 deg beamwidths

Fig. 7 Throughput for 180 deg beamwidths

### B. Optimum bound and concurrency gain

From the above results, it is clear that a small antenna beamwidth provides a better throughput for all schemes. Therefore, a network throughput comparison of MHCT, EMHCT-F, EMHCT-E, and the optimum results of the water-filling solution, as shown in Fig. 8, was conducted for a 20 deg beamwidth. EMHCT-F has highest concurrency gain ($\rho$) as compare to MHCT and EMHCT-E therefore, to obtain the upper bound of the optimum result for the water-filling solution, we used the concurrency gain ($\rho$) of EMHCT-F. It is clear that EMHCT-F and EMHCT-E are better sub-optimum solutions compared to MHCT, because both schemes provide more compact time allocation map compare to MHCT.

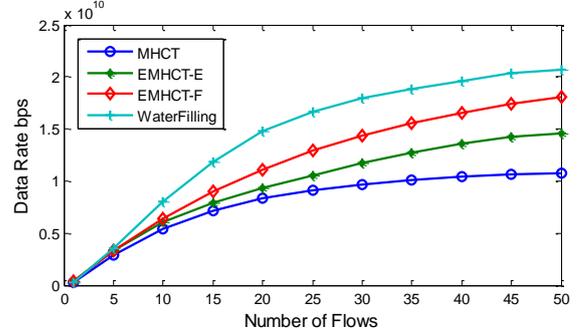

Fig. 8. Throughput of MHCT, EMHCT-F/E, and water-filling

The concurrency gains ($\rho$) of MHCT, EMHCT-F, and EMHCT-E are shown in Fig. 9. EMHCT-F and EMHCT-E achieve a higher concurrency gain compared to MHCT. It is obvious, EMHCT-E and EMHCT-F have more compact time allocation map compare to MHCT, which allow higher number of transmissions to be schedule concurrently. MHCT attain a constant concurrency gain ($\rho$) after 20 flows while EMHCT-E and EMHCT-F approaches to constant concurrency gain ($\rho$) after 45 flows. We call it saturation point of scheduling algorithm. Once saturation occurs the network throughput also approach to a constant value.

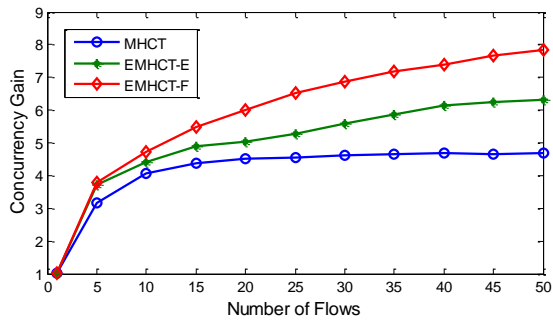

Fig. 9. Concurrency gain (ρ) of MHCT and EMHCT-F/E

*C. Flow throughput fairness*

EMHCT-F and EMHCT-E has higher fairness compare to MHCT. In EMHCT-F and EMHCT-E, each group holds more hop transmission requests compared to MHCT, and hence we get more compact and dense time allocation map. The size of each group will also be nearly the same because the size of a group is determined based on the priority scheme, which is the same for MHCT, EMHCT-F and EMHCT-E. This implies majority of hop transmissions gets same time allocation i-e, $n(G_i)$, hence fairness increases with more compact time allocation.

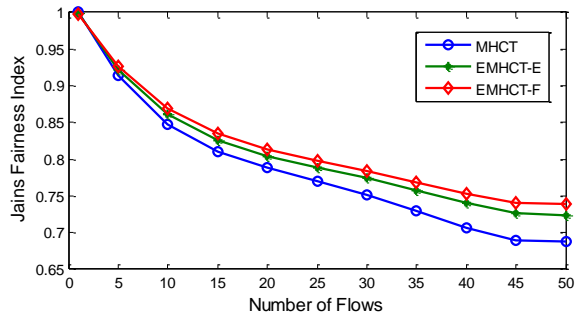

Fig. 10. Fairness of MHCT vs. EMHCT-F/E

4. Algorithm complexity of MHCT and EMHCT-F/E

We took the worst case scenarios to determine the complexity of the algorithms. Let us consider $N$ number of traffic flows with a maximum of $P$ number of hops in a path. A path can have a maximum of $P= n-1$ hops, where $n$ is the total number of nodes. PNC therefore has to perform a sorting (merge sorting with maximum computational time of $NlogN$) of $N$ elements for $N*(n-1)$ number of times, and it has to make $N$ number of comparisons for $N*(n-1)$ number of times. This means that, under the worst condition, to schedule a one-hop transmission, MHCT requires a computational time of $Nlog_2N+N+1$. To schedule all hop requests, the total computational time is $N*(n-1)*(Nlog_2N+N+1)$. If $n-1$ is kept constant, the computational complexity of MHCT is $O(N^2)$.

The method to determine the worst case complexity of EMHCT-F/E is the same as for MHCT, except that for the scheduling of each hop, the number of comparisons under a worst case scenario is $N*(n-1)$. This means that scheduling a one-hop transmission for EMHCT-F/E requires $Nlog_2N+(N*(n-1))+1$ computational time steps. To schedule all hop requests, the total computational time is $N*(n-1)*(Nlog_2N+(N*(n-1))+1)$. If $n-1$ is kept constant, the computational complexity of EMHCT-F/E is also $O(N^2)$.

## V. Conclusion

This paper analyzed the process of a multi-hop concurrent transmission for mmWave communication, considering WPAN in a single room. Based on the analysis of the proposed algorithm, a margin of improvement was found when we considered the relationship of collisions between hop transmissions in the concurrent groups. Thus, for better bandwidth efficiency, we proposed two enhanced schemes of group span overlapping to reduce the total number of allocated time slots during a transmission period for the given transmission requests. In addition, we explicitly showed through a simulation that span overlapping is beneficial. The performances of MHCT, EMHCT-E, and EMHCT-F were also compared with the water-filling solution. From the performance comparison of EMHCT-F/E and the ideal curve of water-filling, it is clear that there is still a possibility for additional improvement. In addition to the further improvement of the scheduling algorithm, the throughput can also be increased through other techniques. For instance, the performance is highly dependent upon the node density in a localized region because a high density leads to a reduction in the average distance between nodes. However, we can predict that the performance will keep increasing until the average distance between nodes approaches the radioactive near field. We assumed that all nodes can transmit with maximum transmission power (P). With high transmission power a hop transmission occupies a large transmission area, causes more interference to other transmissions, and hence reduces the probability of concurrent transmission. With optimum transmission power allocation the probability of concurrent transmission can further be increased.

## References

[1] IEEE 802.15 WPAN Millimeter Wave Alternative PHY Task Group 3c (TG3c). http://www.ieee802.org/15/pub/TG3c.html.

[2] IEEE 802.11 VHT Study Group. Available: http://www.ieee802.org/11/Reports/vht_update.htm.

[3] J. Lee, Y. Chen, and Y. Huang, "A Low-Power Low-Cost Fully-Integrated 60-GHz Transceiver System With OOK Modulation and On-Board Antenna Assembly," IEEE Journal of Solid–State Circuits, vol. 45, no. 2, Feb. 2010.

[4] L. X. Cai *et al.*, "Efficient Resource Management for mmWave WPANs," in Proc. WCNC 2007, pp. 3819-3824.

[5] M. Park and P. Gopalakrishnan," Analysis on Spatial Reuse and Interference in 60-GHz Wireless Networks," IEEE Journal on Selected Areas in Communications, vol. 27, no. 8, Oct. 2009.

[6] F. Yildirim and H. Liu, "A Cross-Layer Neighbor-Discovery Algorithm for Directional 60-GHz Networks," IEEE Trans. Veh. Technol., vol. 58, no. 8, Oct. 2009.

[7] M. Park *et al.*, "Millimeter-Wave Multi-Gigabit WLAN: Challenges and Feasibility," in IEEE 19th International Symposium on Personal, Indoor and Mobile Radio Communications, 2008.

[8] S. Singh *et al.*, "Millimeter Wave WPAN: Cross-Layer Modeling and Multihop Architecture," in Proc. IEEE INFOCOM'07, May 2007, pp.2336-2240.

[9] L. X. Cai *et al.*, "REX: a Randomized EXclusive Region based Scheduling Scheme for mmWave WPANs with Directional Antenna," IEEE Trans. Wireless Commun., vol. 9, no. 1, pp. 113-121, 2010.

[10] L. X. Cai *et al.*, "Spatial Multiplexing Capacity Analysis of mmWave WPANs with Directional Antenna," in Proc.IEEE GLOBECOM'07, Novermber 2007, pp. 4744-4748.

[11] J. Wang, R.Venkatesha Prasad, and I.G.M.M. Niemegeers, "Enabling Multihop on mm Wave WPANs," in IEEE ISWCS'08, Oct. 2008, pp. 371-375.

[12] J. Qiao, L.X. Cai, and X. Shen, "Multi-Hop Concurrent Transmission in Millimeter Wave WPANs with Directional Antenna," in Proc. IEEE ICC'10, May 2010, pp.1-5.

[13] R. Mudumbai, S. Singh, and U. Madhow, "Medium Access Control for 60 GHz Outdoor Mesh Networks with Highly Directional Links," in Proc.IEEE INFOCOM'09, April 2009, pp. 2871-2875...

[14] S. Y. Geng *et al.*, "Millimeter-Wave Propagation Channel Characterization for Short-Range Wireless Communications," IEEE Trans. Veh. Technol., vol. 58, no. 1, pp.3-13, Jan. 2009.

[15] Z. Yang *et al.*, "Practical Scheduling Algorithms for Concurrent Transmissions in Rate-adaptive Wireless Networks," in Proc. IEEE INFOCOM'10, March 2010.

[16] S. Collonge, G. Zaharia, and G. El Zein, "Influence of the Human Activity on the Propagation Characteristics of 60 GHz Indoor Channels," IEEE Transactions on Wireless Communications, vol. 3, no. 6, Nov. 2004.

[17] L. X. Cai *et al.*, "REX: A Randomized EXclusive Region Based Scheduling Scheme for mmWave WPANs with Directional Antenna," IEEE Transactions on Wireless Communications, vol. 9, no. 1, Jan. 2010.

[18] D. P. Palomarand and J. R. Fonollosa, "Practical Algorithms for a Family of Waterfilling Solutions," IEEE Transactions on Signal Processing, vol. 53, no. 2, Feb. 2005.

[19] J. Mo and J. Walrand, "Fair End-to-End Window-Based Congestion Control," IEEE/ACM Transactions on Networking, vol. 8, no. 5, Oct. 2000.